\documentclass[3p,times]{elsarticle}

\usepackage{amssymb}
\usepackage{amsxtra}
\usepackage{amsmath}
\usepackage{amsfonts}
\usepackage{CJK}
\usepackage[titletoc]{appendix}
\usepackage{graphicx}
\usepackage{color}

\numberwithin{equation}{section}

\newcommand{\eqa}{\begin{eqnarray}}
\newcommand{\eeqa}{\end{eqnarray}}
\newcommand{\beq}{\begin{equation}}
\newcommand{\eeq}{\end{equation}}
\newcommand{\nn}{\nonumber}

\allowdisplaybreaks

\begin{document}
\begin{frontmatter}
\title{Riemann-Hilbert method and N-soliton solutions for the mixed Chen-Lee-Liu derivative nonlinear Schr\"{o}dinger equation}
\author{Fang Fang}
\author{Beibei Hu\corref{cor1}}
\author{Ling Zhang}

\cortext[cor1]{Corresponding author. Email: hu\_chzu@163.com; hu\_chzu@shu.edu.cn}

\address{School of Mathematics and Finance, Chuzhou University, Anhui 239000 China}

\pagestyle{plain}
\setcounter{page}{1}
\begin{abstract}
In this paper, we aim to investigate the mixed Chen-Lee-Liu derivative nonlinear Schr\"{o}dinger(CLL-NLS) equation via the Riemann-Hilbert(RH) method. we construct a RH problem base on the Jost solution of the Lax pair. By solving this RH problem corresponding to the non reflection case, the N-soliton solution of CLL-NLS equation is obtained, which expression is the ratio of $(2N+1)\times(2N+1)$ determinant and $2N\times2N$ determinant.
\end{abstract}
\end{frontmatter}

\section{Introduction}
Soliton theory is an important branch of nonlinear science. In physics, solitons are used to describe solitary waves with elastic scattering characteristics. In mathematics, soliton theory provides a series of methods for solving nonlinear partial differential equations, which attracts the attention of mathematical physicists. With the development of soliton theory, many methods to solve soliton equations with important physical background have been proposed, for example, inverse scattering transform(IST) \cite{1,2,3}, Darboux transformation(DT) \cite{8}, B\"{a}cklund transformation \cite{9}, Hirota bilinear method \cite{10,11,12}, Wroskian technique \cite{14,15,16} and so on.

The derivative nonlinear schr\"{o}dinger(DNLS) equation describing Alfv\'{e}n waves in magnetic field is as follows \cite{17}:
\beq iu_{t}=u_{xx}+i(|u|^2u)_{x}=0, \label{1.1}\eeq
which is one of the most significant equations in physics. Kaup and Newell obtained the solution of Eq.\eqref{1.1} by using IST method\cite{18}. The IST method has obvious advantages in solving the initial value of the soliton equation, but its calculation is large. Fortunately, on the basis of IST method, Riemann-Hilbert(RH) method, a relatively simple and direct method for solving soliton equation, was proposed by Novikovet et al.\cite{19}. This method is similar to the IST method, the RH method first considers the direct scattering problem, that is, the RH problem is constructed, from the initial data of the soliton equation, the scattering data at the initial time is obtained, and the scattering data at any time is obtained by using its time evolution law. Finally, the exact solution is established by using the IST method\cite{20}. Since the RH method was proposed, many solutions of soliton equations have been discussed, such as, the coupled DNLS equation \cite{21}, the coupled higher-order NLS equation \cite{22}, the short pulse(SP) equation \cite{23}, the coupled modified Korteweg-de Vries (mKdV) equation \cite{24}, the generalized Sasa-Satsuma equation \cite{25}, the two-component Gerdjikov-Ivanov(GI) equation \cite{26}, the modified SP equation \cite{27}, and so on \cite{28,29,30}. In particular, RH method is an effective way to working the initial-boundary value problems(IBVPs) of the integrable systems \cite{31,32,33,34,35,36}.

Recently, Chan et al.\cite{37} reported the the mixed Chen-Lee-Liu derivative nonlinear Sch\"{o}dinger(CLL-NLS) equation as follows
\beq ir_{t}+r_{xx}+|r|^2r-i|r|^2r_{x}=0, \label{1.2}\eeq
which is a completely integrable model, and a large number of solutions of Eq.\eqref{1.2} are discussed, such as, the soliton solution by Hirota bilinear method \cite{38}, the higher-order soliton, breathers, and rogue wave solutions by DT method\cite{39}. In particular, the IBVPs of Eq.\eqref{1.2} to be investigated by Fokas method\cite{40}.

The design structure of this paper is as follows. Section 2, we will construct a basic RH problem based on the Jost solution of Lax pair. Section 3, we will give the reconstruction of potential function and the law of scattering data evolution with time. Section 4, the formula of N-soliton solution expressed by determinant ratio is proposed. Section 5 is the conclusions.

\section{The Riemann-Hilbert problem}

In this section, we shall construct a RH problem for the CLL-NLS equation, by using IST method. First of all, we introduce the coupled CLL-NLS equation as following:
\beq\left\{\begin{array}{l}
r_{t}-ir_{xx}+ir^2q+rqr_{x}=0,\\
q_{t}+iq_{xx}-iq^2r+qrq_{x}=0.
\end{array}\right.\label{2.1}\eeq
which reduces to the CLL-NLS equation while $q=-r^{*}$ and the $*$ denotes complex conjugation. These two equations in \eqref{2.1} are the compatibility condition of the following Lax pair\cite{39,40,42,43}
\begin{subequations}
\begin{align}
&\Phi_x=U\Phi=\left(i (\lambda^{2}-\frac{1}{2})\sigma+\lambda Q+\frac{1}{4}i Q^{2}\sigma\right)\Phi,\label{2.2a}\\
&\Phi_t=V\Phi=[
-2i(\lambda^2-\frac{1}{2})^{2}\sigma-2\lambda^3Q-i\lambda^2Q^2\sigma+\lambda\left(Q+i\sigma Q_x-\frac{1}{2}Q^3\right)\nn\\
&\qquad\qquad\quad-\frac{1}{8}iQ^4\sigma+\frac{1}{4}(QQ_x-Q_xQ)]\Phi,\label{2.2b}
\end{align}
\end{subequations}
with
\eqa\begin{array}{l}
\sigma=\left(\begin{array}{cc}
1&0\\
0&-1\end{array} \right),
Q=\left(\begin{array}{cc}
0&r\\
-r^{*}&0\end{array} \right).
\end{array}\label{2.3}\eeqa
where $\Phi=\Phi(x,t;\lambda)$ is a matrix function of the complex spectral parameter, $\lambda$ is the spectral parameter. $Q$ is named potential function.

For the sake of convenience, we introduce a new matrix function $J=J(x,t;\lambda)$ defined by
\beq \Phi=Je^{i(\lambda^2-\frac{1}{2})\sigma x-2i(\lambda^2-\frac{1}{2})^{2}\sigma t}\label{2.4}\eeq
Obviously, we can check that $Je^{i(\lambda^2-\frac{1}{2})\sigma x-2i(\lambda^2-\frac{1}{2})^{2}\sigma t}$ satisfies Eqs.\eqref{2.2a},\eqref{2.2b}. Inserting \eqref{2.4} into \eqref{2.2a}-\eqref{2.2b}, the form of the Lax pair \eqref{2.2a}-\eqref{2.2b} becomes
\begin{subequations}
\begin{align}
&J_x=i(\lambda^2-\frac{1}{2})[\sigma,J]+U_{1}J,\label{2.5a}\\
&J_t=-2i(\lambda^2-\frac{1}{2})^{2}[\sigma,J]+V_1J.\label{2.5b}
\end{align}
\end{subequations}
where $[\sigma,J]=\sigma J-J\sigma$ is the commutator.
\begin{subequations}
\begin{align}
&U_{1}=\lambda Q+\frac{1}{4}iQ^{2}\sigma
&V_1=-2\lambda^3Q-i\lambda^2Q^2\sigma+\lambda\left(Q+i\sigma Q_x-\frac{1}{2}Q^3\right)-\frac{1}{8}iQ^4\sigma+\frac{1}{4}(QQ_x-Q_xQ)
\nn\end{align}
\end{subequations}

As usual, in the following scattering process, we only concentrate on the $x$-part of the Lax pair \eqref{2.2a}. Indeed the $x$-part of the Lax pair \eqref{2.2a} allows us to take use of the existing symmetry relations of the potential Q. Consequently, we shall treat the time t as a dummy variable and omit it. Now, we calculate two Jost solutions $J_{\pm}= J_{\pm}(x,\lambda)$ of Eq.\eqref{2.5a} for $\lambda\in \mathbb{R}$. Based on the properties of the Jost solutions $J_{\pm}= J_{\pm}(x,\lambda)$
\eqa J_{+}=([J_{+}]_1,[J_{+}]_2),\label{2.6}\eeqa
\eqa J_{-}=([J_{-}]_1,[J_{-}]_2),\label{2.7}\eeqa
with the boundary conditions
\begin{subequations}
\begin{align}
&J_{+}\rightarrow \mathrm{I},\,\,x\rightarrow -\infty,\label{2.8a}\\
&J_{-}\rightarrow \mathrm{I},\,\,x\rightarrow +\infty.\label{2.8b}
\end{align}
\end{subequations}
the subscripts in $J(x,\lambda)$ represent which end of the $x$-axis the boundary
conditions are set. Where $[J_{\pm}]_n(n=1,2)$ denote the $n$-th column vector of $J_{\pm}$, $\mathrm{I}=diag\{1,1\}$ is the $2\times2$ unit matrix.
According to the method of variation of parameters as well as the boundary conditions, we can turn Eq.\eqref{2.5a} for $\lambda\in \mathbb{R}$ into the Volterra integral equations.
\eqa J_{+}(x,\lambda)=\mathrm{I}-\int_x^{+\infty}e^{i(\lambda^2-\frac{1}{2})\hat\sigma (x-\xi)}U_{1}J_{+}(\xi,\lambda)d\xi,\label{2.9}\eeqa
\eqa J_{-}(x,\lambda)=\mathrm{I}+\int_{-\infty}^xe^{i(\lambda^2-\frac{1}{2})\hat\sigma (x-\xi)}U_{1}J_{-}(\xi,\lambda)d\xi,\label{2.10}\eeqa
where $\hat\sigma$ represents a matrix operator acting on $2\times2$ matrix $X$ by $\hat\sigma X=[\sigma,X]$ and by
$e^{x\hat\sigma}X=e^{x\sigma}Xe^{-x\sigma}$. Moreover, after simple analysis, we find that $[J_{+}]_1, [J_{-}]_2$ are analytic for $\lambda\in D_{+}$ and continuous for $\lambda\in D_{+}\bigcup\mathbb{R}\bigcup i\mathbb{R}$. Similarly $[J_{-}]_1, [J_{+}]_2$ are analytic for $\lambda\in D_{-}$ and continuous for $\lambda\in D_{-}\bigcup\mathbb{R}\bigcup i\mathbb{R}$,
here
\begin{subequations}
\begin{align}
&D_{+}=\left\{\lambda|arg\lambda\in\left(0,\frac{\pi}{2}\right)\bigcup\left(\pi,\frac{3\pi}{2}\right)\right\}\nn\\
&D_{-}=\left\{\lambda|arg\lambda\in\left(\frac{\pi}{2},\pi\right)\bigcup\left(\frac{3\pi}{2},2\pi\right)\right\}.\nn
\end{align}
\end{subequations}

Secondly, let us investigate the properties of ${J}_{\pm}$. We deduce the determinants of ${J}_{\pm}$ are constants for all $x$ on the basis of the Abel's identity and $\mathrm{Tr}(U_{1})=0$. Furthermore, due to the boundary conditions Eq.\eqref{2.8a},\eqref{2.8b}, we have
\beq \det {{J}_{\pm }}=1,\quad \lambda \in \mathbb{R}\bigcup i\mathbb{R}. \label{2.11}\eeq
By introducing a new function $ E(x,\lambda)=e^{i(\lambda^2-\frac{1}{2})\sigma x}$, we find that spectral problem Eq.\eqref{2.5a} exists two fundamental matrix solutions ${{J}_{+}}E$ and ${{J}_{-}}E$, which are linearly related by a $2\times2$ scattering matrix $S(\lambda)$
\beq {{J}_{-}}E={{J}_{+}}E\cdot S(\lambda),\quad \lambda \in \mathbb{R}\bigcup i\mathbb{R}.\label{2.12}\eeq
From Eq.\eqref{2.11} and \eqref{2.12}, we know that
\beq \det S(\lambda)=1. \label{2.14}\eeq
Furthermore, let $x$ go to $+\infty$, the $2\times2$ scattering matrix $S(\lambda)$ is given as
\beq  S(\lambda)=\lim_{x\rightarrow+\infty}E^{-1}J_{-}E=\mathrm{I}+\int_{-\infty}^{+\infty}E^{-1}U_{1}J_{-}Ed\xi,\,\,\lambda \in \mathbb{R}\bigcup i\mathbb{R}.\nn\eeq
From the analytic property of $J_{-}$, we find that $s_{22}$ can be analytically extended to $D_{+}$, $s_{11}$
allows analytic extensions to $D_{-}$. Generally speaking, $s_{12}$, $s_{21}$ can only be defined in the $\mathbb{R}\bigcup i\mathbb{R}$.

In what follows, we shall construct a RH problem for the CLL-NLS equation. Firstly, using the analytic properties of $J_{\pm}$, we define a new Jost solution $P_1=P_1(x,\lambda)$ as
\beq P_{1}=([J_{+}]_1,[J_{-}]_2).\label{2.13}\eeq
which is obviously analytic for $\lambda\in D_{+}$. In addition, from Eqs. \eqref{2.9} and \eqref{2.10}, we have
\beq P_{1}\rightarrow\mathrm{I},\qquad\qquad\qquad\lambda \rightarrow +\infty,\, \lambda\in D_{+}. \label{2.20}\eeq
Next, we introduce the limit of $P_{1}$
\beq
P^{+}=\lim_{\lambda\to\Gamma}P_1\qquad\qquad\qquad\Gamma=\mathbb{R}\bigcup i\mathbb{R}.\eeq
From \eqref{2.12} and \eqref{2.13}, we can get
\beq
P^{+}=J_{+}\left(\begin{array}{cc}
1&e^{i(2\lambda^2-1)x}s_{12}\\
0&s_{22}\
\end{array}\right).\label{2.26}\eeq

To obtain the analytic counterpart of $P^{+}$ in $D_{-}$, denoted by $P_{2}$, we consider the inverse matrices $J_{\pm}^{-1}$ defined as
\beq J_{+}^{-1}=\left(\begin{array}{cc}
{[J_{+}^{-1}]}_1\\
{[J_{+}^{-1}]}_2
\end{array}\right),\quad
{J_{-}^{-1}}=\left(\begin{array}{cc}
{[J_{-}^{-1}]}_1\\
{[J_{-}^{-1}]}_2
\end{array}\right),\label{2.22}\eeq
 here $[J_{\pm}^{-1}]_n (n=1,2) $ denote the $n$-th row vector of $J_{\pm}^{-1}$. Then we can see that ${[J_{+}^{-1}]}_1, {[J_{-}^{-1}]}_2$ are analytic for $\lambda\in D_{-}$ and continuous for $\lambda\in D_{-}\bigcup\mathbb{R}\bigcup i\mathbb{R}$, whereas ${[J_{-}^{-1}]}_1$, ${[J_{+}^{-1}]}_2$ are analytic for $\lambda\in D_{+}$ and continuous for $\lambda\in D_{+}\bigcup\mathbb{R}\bigcup i\mathbb{R}$. Obviously, $J_{\pm}^{-1}$ satisfy the adjoint scattering equation of Eq.\eqref{2.5a}:
\beq K_x=i(\lambda^2-\frac{1}{2})[\sigma,K]-KU_{1}. \label{2.21}\eeq
In addition, it is not difficult to find that the inverse matrices $J_{+}^{-1}$ and $J_{-}^{-1}$ satisfy the following boundary conditions.
\begin{subequations}
\begin{align}
&J_{+}^{-1}\rightarrow \mathrm{I},\,\,x\rightarrow -\infty,\label{2.23a}\\
&J_{-}^{-1}\rightarrow \mathrm{I},\,\,x\rightarrow +\infty.\label{2.23b}
\end{align}
\end{subequations}
Taking the similar procedure as above, a matrix function $P_{2}$ which is analytic in $D_{-}$
\beq P_{2}=\left(\begin{array}{cc}
{[J_{+}^{-1}]}_1\\
{[J_{-}^{-1}]}_2
\end{array}\right).\label{2.24}\eeq
and $P^{-}$
\beq
P^{-}=\lim_{\lambda\to\Gamma}P_2\qquad\qquad\qquad\Gamma=\mathbb{R}\bigcup i\mathbb{R}.\eeq
are expressed. Moreover, we can get that
\beq P_{2}\rightarrow\mathrm{I},\quad\quad\quad \lambda\rightarrow -\infty,\, \lambda\in D_{-}. \label{2.25}\eeq
and
\beq
P^{-}=\left(\begin{array}{cc}
1&0\\
e^{-i(2\lambda-1)x}r_{21}&r_{22}
\end{array}\right)J_{+}^{-1}.\label{2.27}\eeq
with $R(\lambda)\equiv (r_{kj})_{2\times2}=S^{-1}(\lambda)$. Similar to the scattering coefficients $s_{ij}$ above, it is easy to know that $r_{11}$ allows an analytic extension to $D_{+}$ and $r_{22}$
is analytically extendible to $D_{-}$. Generally speaking, $r_{12}$, $r_{21}$ can only be defined in the $\mathbb{R}\bigcup i\mathbb{R}$. In addition, from \eqref{2.12}, we get
\beq
E^{-1}J_{-}^{-1}=R(\lambda)\cdot E^{-1}J_{+}^{-1},\quad\quad\lambda\in\mathbb{R}\bigcup i\mathbb{R}.\eeq

Summarizing the above the results, we find that two matrix functions $P^{+}$ and $P^{-}$ which are analytic in $D_{+}$ and $D_{-}$, respectively, are related by
\beq P^{-}(x,\lambda)P^{+}(x,\lambda)=\left(\begin{array}{cc}
1 & s_{12}e^{i(2\lambda-1)x}\\
r_{21}e^{-i(2\lambda-1)x}& 1
\end{array}\right),\,\, \lambda\in \mathbb{R}\bigcup i\mathbb{R}. \label{2.29}\eeq
Eq.\eqref{2.29} is just the RH problem for the CLL-NLS equation. In order to obtain solution for this RH problem, we assume that the RH problem is non-regular when $\mathrm{det}P_1$ and $\mathrm{det}P_2$  can be zero for $\lambda_k\in D_{+}$ and $\lambda_k\in D_{-}$, respectively. $1\leq k\leq N$. where $N$ is the number of these zeros. Recalling the definitions of $P_{1}$ and $P_{2}$, we can see that
\eqa
\mathrm{det} P_{1}(x,\lambda)=s_{22}(\lambda),\quad\quad\quad\lambda\in D_{+},\\
\mathrm{det} P_{2}(x,\lambda)=r_{22}(\lambda),\quad\quad\quad\lambda\in D_{-}.
\eeqa
To specify these zeros, we first can take use of a symmetry relation for $ U_{1}$
$$U_{1}^\dag=\sigma U_{1} \sigma,$$
where the superscript $\dag$ means the Hermitian of a matrix. Therefore from Eq. \eqref{2.21}, we arrive at
\eqa
J_{\pm}^{\dag}(x,t,\lambda^*)=\sigma J_{\pm}^{-1}(x,t,\lambda)\sigma,\label{3.2}\eeqa
Then from \eqref{2.12}, we also gain
\eqa
S^{\dag}(\lambda^*)=\sigma S^{-1}(\lambda)\sigma,\label{3.3}\eeqa
which implies the following relations
\eqa
&&r_{11}(\lambda)=s^{*}_{11}(\lambda^*)\quad\quad\lambda\in D_{+},\\
&&r_{22}(\lambda)=s^{*}_{22}(\lambda^*)\quad\quad\lambda\in D_{-},\label{3.14}\\
&&r_{12}(\lambda)=-s^{*}_{21}(\lambda^*)\quad\quad\lambda\in\mathbb{R}\bigcup i\mathbb{R},\\
&&r_{21}(\lambda)=-s^{*}_{12}(\lambda^*)\quad\quad\lambda\in\mathbb{R}\bigcup i\mathbb{R}.\eeqa
Moreover from Eq. \eqref{3.3} and the definitions of $P_1$, $P_2$, we point out that the analytic solutions $P_1$, $P_2$ satisfy the involution property
\beq P_{1}^{\dag}(x,\lambda^*)=\sigma P_{2}(x,\lambda)\sigma,\quad\quad\quad\lambda\in D_{-}\label{3.4}\eeq

The similar analysis shows that the potential matrix $Q$ also satisfies another symmetry relations
$$Q=-\sigma Q\sigma$$
It follows that
\beq J_{\pm}(-\lambda)=\sigma J_{\pm}(\lambda)\sigma.\eeq
and
\beq P_1(-\lambda)=\sigma P_1(\lambda)\sigma.\eeq
thus
\eqa
&&s_{11}(-\lambda)=-s_{11}(\lambda)\quad\quad\lambda\in D_{-},\\
&&s_{22}(-\lambda)=-s_{22}(\lambda)\quad\quad\lambda\in D_{+},\label{3.15}\\
&&s_{12}(-\lambda)=-s_{12}(\lambda)\quad\quad\lambda\in\mathbb{R}\bigcup i\mathbb{R},\\
&&s_{21}(-\lambda)=-s_{21}(\lambda)\quad\quad\lambda\in\mathbb{R}\bigcup i\mathbb{R}.\eeqa

Therefore, from \eqref{3.14}, we find that if $\lambda_{j}$ is a zero of det$P_{1}$, then $\hat\lambda_{j}=\lambda^{*}_{j}$ is a zero of det$P_{2}$. Moreover, in view of \eqref{3.15}, we know that $-\lambda$ is also a zero of det$P_1$. Hence we suppose that det$P_1$ has $2N$ simple zeros $\{\lambda_j\}_1^{2N}$ satisfying $\lambda_{N+l}=-\lambda_l\quad(1\leq l \leq N)$, which all lie in $D_+$. From the zeros of det$P_1$, we see that det$P_2$ possesses $2N$ simple zeros $\{\hat{\lambda_j}\}_1^{2N}$ satisfying $\hat{\lambda_j}=\lambda_j^*,\quad(1\leq j\leq2N)$, which are all in $D_-$. Obviously the zeros of det$P_1$ and det$P_2$ always appear in quadruples. To solve the RH problem, we need the scattering data including the continuous scattering data $\{s_{12},s_{21}\}$ and the discrete scattering data $\{\lambda_j,\hat\lambda_j,v_j,\hat v_j\}$ which a single column vector $v_{j}$ and row vector $\hat v_{j}$ satisfying
\eqa P_{1}(\lambda_j)v_{j}=0,\quad \hat v_{j}P_{2}(\hat \lambda_j)=0.  \label{3.4}\eeqa

On the one hand, by taking the Hermitian  conjugate of $P_{1}(\lambda_j)v_{j}=0$, we can construct the relationship between each pair of $v_j$ and $\hat v_j$.
\eqa
&&v_{j}=\sigma v_{j-N}\quad\quad\quad N+1\leq j\leq2N,\nn\\
&&\hat v_j=v_j^{\dag}\sigma\quad\quad\quad 1\leq j\leq N.\label{3.8}\eeqa

On the other hand, in order to obtain the spatial evolutions for vectors $v_j(x)$, taking the $x$-derivative to equation $P_{1}v_j=0$ and using \eqref{2.5a}, we obtain
\beq v_{j}=e^{i(\lambda_j^2-\frac{1}{2})\sigma x}v_{j_{0}}\quad\quad\quad1\leq j\leq N,\label{3.9}\eeq
where $v_{j_{0}}=v_j|_x=0$.

Using these vectors, the RH problem which corresponds to the reflection-less case, that is to say, we set the vanishing cofficient $s_{12}=s_{21}=0$ in the RH problem \eqref{2.29}, can be solved exactly,and the result is
\begin{subequations}
\begin{align}
&P_{1}(\lambda)=\mathrm{I}-\sum_{k=1}^{2N}\sum_{j=1}^{2N}\frac{v_k(M^{-1})_{kj}\hat v_j}{\lambda-\hat\lambda_j},\label{4.1a}\\
&P_{2}(\lambda)=\mathrm{I}+\sum_{k=1}^{2N}\sum_{j=1}^{2N}\frac{v_k(M^{-1})_{kj}\hat v_j}{\lambda-\lambda_k}.\label{4.1b}
\end{align}
\end{subequations}
where $M=(m_{kj})_{2N\times 2N}$ is a matrix whose entries are
\eqa m_{kj}=\frac{\hat v_kv_j}{\lambda_j-\hat\lambda_k},\,1\leq k,j\leq 2N.\label{4.2}\eeqa
\section{ Inverse scattering transform }

In this section, with the help of $P_{1}$ in \eqref{4.1a}, we can write out explicitly the potential $Q$. Owing to $P_{1}(\lambda)$ is the solution of spectral problem \eqref{2.5a}, we assume that the asymptotic expansion of $P_{1}(\lambda)$ at large $\lambda$ as
\beq P_{1}=\mathrm{I}+\frac{P_{1}^{(1)}}{\lambda}+\frac{P_{1}^{(2)}}{\lambda^{2}}+O(\lambda^{-3})\quad \lambda\rightarrow\infty,\label{3.5}\eeq
Then by substituting the above expansion into (2.5a) and comparing $O(\lambda)$ terms,we obtains
\beq Q=-i[\sigma,P_{1}^{(1)}]=\left(\begin{array}{cc}
0 & -2i(P_{1}^{(1)})_{12}\\
2i(P_{1}^{(1)})_{21} & 0
\end{array}\right).\label{3.6}\eeq
which implies that $r$ can be reconstructed as
\beq r=-2i(P_{1}^{(1)})_{12}.\label{3.7}\eeq
where $P_{1}^{(1)}=(P_{1}^{(1)})_{2\times2}$ and $(P_{1}^{(1)})_{ij}$ is the $(i;j)$-entry of $P_{1}^{(1)},i,j=1,2$.
Here, the matrix function $P_{1}^{(1)}$ can be found from \eqref{4.1a}
\beq P_{1}^{(1)}=\sum_{k=1}^{2N}\sum_{j=1}^{2N}v_k(M^{-1})_{kj}\hat v_j.\label{4.3}\eeq

\section{The soliton solutions}

To derive the solutions for the CLL-NLS equation, we also need the scattering data at time $t$, which need investigate the time evolution of scattering data. In fact, by using \eqref{2.5b} and \eqref{2.12}, making the limit $x\rightarrow+\infty$, and taking into account the boundary condition \eqref{2.8a} for $J_{+}$ as well as $V_1\rightarrow0$ as $x\rightarrow\pm\infty$, we arrive at
\eqa
&&s_{11}|_t=s_{22}|_t=0,\quad s_{12}|_t=-4i(\lambda^2-\frac{1}{2})^{2} s_{12},\quad s_{21}|_t=-4i(\lambda^2-\frac{1}{2})^{2}s_{21},\nn\\
&&\frac{d\lambda_{j}}{dt}=0,\quad v_j|_t=-2i(\lambda^2-\frac{1}{2})^{2} v_j.\label{3.16}\eeqa
Combining \eqref{3.8} with \eqref{3.9}, we can derive the column vectors $v_j$ and the row vector $\hat v_j$ explicitly,
\beq
v_j=\left\{\begin{array}{l}
e^{\theta_j\sigma}v_{j_{0}},\quad\quad\quad\quad 1\leq j\leq N\\
\sigma e^{\theta_{j-N}\sigma}v_{j-N,0},\quad N+1\leq j\leq 2N
\end{array}\right.\label{5.1}\eeq
and
\beq
\hat v_j=\left\{\begin{array}{l}
v_{j_{0}}^{\dag}e^{\theta_j^*\sigma}\sigma\quad\quad\quad 1\leq j\leq N\\
v_{j-N,0}^\dag e^{\theta_{j-N}^*\sigma},\quad N+1\leq j\leq 2N
\end{array}\right.\label{5.2}\eeq
where $\theta_j=i(\lambda_j^2-\frac{1}{2})x-2i(\lambda_{j}^2-\frac{1}{2})^{2}t\quad(\lambda_j\in D+)$, $v_{j_{0}}$ is a constant vector.

we have chosen $v_{j_{0}}=(c_j,1)^T$, it follows from \eqref{4.3} that the N-soliton solutions for the CLL-NLS equation reads
\eqa
&&r=2i\sum_{k=1}^{N}\sum_{j=1}^{N}c_ke^{\theta_k-\theta_j^*}(M^{-1})_{kj}+2i\sum_{k=N+1}^{2N}\sum_{j=1}^{N}c_{k-N}e^{\theta_{k-N}-\theta_j^*}(M^{-1})_{kj}\nn\\
&&-2i\sum_{k=1}^{N}\sum_{j=N+1}^{2N}c_ke^{\theta_k-\theta_{j-N}^*}(M^{-1})_{kj}-2i\sum_{k=N+1}^{2N}\sum_{j=N+1}^{2N}c_{k-N}e^{\theta_{k-N}-\theta_{j-N}^*}(M^{-1})_{kj}.
\label{5.3}\eeqa
and $M=(m_{kj})_{2N\times2N}$ is given by
\eqa
m_{kj}=\left\{\begin{array}{l}
\frac{c_jc_k^*e^{\theta_k^*+\theta_j}-e^{-(\theta_k^*+\theta_j)}}{\lambda_k^*-\lambda_j},\quad\;1\leq k,j\leq N,
\\
\frac{c_{j-N}c_k^*e^{\theta_k^*+\theta_{j-N}}+e^{-(\theta_k^*+\theta_{j-N})}}{\lambda_k^*-\lambda_j},\quad\;1\leq k\leq N,N+1\leq j\leq2N,\\
\frac{c_{j}c_{k-N}^*e^{\theta_{k-N}^*+\theta_{j}}+e^{-(\theta_{k-N}^*+\theta_{j})}}{-\lambda_{k-N}^*-\lambda_j},\quad\;1\leq j\leq N,N+1\leq k\leq2N,\\
\frac{c_{j-N}c_{k-N}^*e^{\theta_{k-N}^*+\theta_{j-N}}-e^{-(\theta_{k-N}^*+\theta_{j-N})}}{-\lambda_{k-N}^*+\lambda_{j-N}},\quad\;N+1\leq k,j\leq2N.
\end{array}\right.\label{5.4}\eeqa
with $\theta_j=i(\lambda_j^2-\frac{1}{2})x-2i(\lambda_{j}^2-\frac{1}{2})^{2}t\quad(\lambda_j\in D+)$

The simplest situation occurs when $N=1$ in formula \eqref{5.3}. the single-soliton solution is
\beq
r=-\frac{2ic_1e^{\theta_1-\theta_1^*}\left(m_{11}+m_{21}-m_{12}-m_{22}\right)}{det M}
\eeq
where $M=(m_{kj})_{2\times2}$ is given by
\eqa
m_{11}=\frac{|c_1|\left(e^{\theta_1^*+\theta_1+\xi_1}-e^{-\theta_1^*-\theta_1-\xi_1}\right)}{\lambda_1^*-\lambda_1},
m_{12}=\frac{|c_1|\left(e^{\theta_1^*+\theta_1+\xi_1}+e^{-\theta_1^*-\theta_1-\xi_1}\right)}{\lambda_1^*+\lambda_1},\nn\\
m_{21}=\frac{|c_1|\left(e^{\theta_1^*+\theta_1+\xi_1}+e^{-\theta_1^*-\theta_1-\xi_1}\right)}{-\lambda_1^*-\lambda_1},
m_{22}=\frac{|c_1|\left(e^{\theta_1^*+\theta_1+\xi_1}-e^{-\theta_1^*-\theta_1-\xi_1}\right)}{-\lambda_1^*+\lambda_1}.\nn\eeqa

Letting $\lambda_1=\lambda_{11}+i\lambda_{12}$, $|c_1|=e^{\xi_1}$,
then the single-soliton solution can be written as
\beq
r=\frac{4\lambda_{11}\lambda_{12}e^{iY}\left(\lambda_{11}sinhX
+i\lambda_{12}coshX\right)}{\lambda_{12}^2cosh^2X+\lambda_{11}^2sinh^2X}.
\label{5.6}\eeq
with
\eqa
&&X=8\lambda_{11}\lambda_{12}(\lambda_{11}^2-\lambda_{12}^2)t-2\lambda_{11}\lambda_{12}x-4\lambda_{11}\lambda_{12}t+ln|c_1|,\nn\\
&&Y=2(\lambda_{11}^2-\lambda_{12}^2)x-4(\lambda_{11}^4-6\lambda_{11}^2\lambda_{12}^2+\lambda_{12}^4-\lambda_{11}^2-\lambda_{12}^2)t.
\nn\eeqa

In order to understand the properties of the resulting soliton solution, we can take the model
\beq
|r|=\frac{4|\lambda_{11}\lambda_{12}|}{\sqrt{\lambda_{12}^2cosh^2X+\lambda_{11}^2sinh^2X}}.
\eeq

Now we investigate the case for $N=2$ in formula \eqref{5.3}. we arrive at the two-soliton solution as follows:
\eqa
&&r=-2i\sum_{k=1}^{2}\sum_{j=1}^{2}c_ke^{\theta_k-\theta_j^*}(M^{-1})_{kj}-2i\sum_{k=3}^{4}\sum_{j=1}^{2}c_{k-2}e^{\theta_{k-2}-\theta_j^*}(M^{-1})_{kj}\nn\\
&&\quad\quad+2i\sum_{k=1}^{2}\sum_{j=3}^{4}c_ke^{\theta_k-\theta_{j-2}^*}(M^{-1})_{kj}+2i\sum_{k=3}^{4}\sum_{j=3}^{4}c_{k-2}e^{\theta_{k-2}-\theta_{j-2}^*}(M^{-1})_{kj}.
\label{5.5}\eeqa
and $M=(m_{kj})_{4\times4}$ is given by
\eqa
m_{11}=\frac{c_1c_1^*e^{\theta_1^*+\theta_1}-e^{-\theta_1^*-\theta_1}}{\lambda_1^*-\lambda_1},
m_{12}=\frac{c_1c_2^*e^{\theta_1^*+\theta_2}-e^{-\theta_1^*-\theta_2}}{\lambda_1^*-\lambda_2},\nn\\
m_{13}=\frac{c_1c_1^*e^{\theta_1^*+\theta_1}+e^{-\theta_1^*-\theta_1}}{\lambda_1^*+\lambda_1},
m_{14}=\frac{c_1^*c_2e^{\theta_1^*+\theta_2}+e^{-\theta_1^*-\theta_2}}{\lambda_1^*+\lambda_2},\nn\\
m_{21}=\frac{c_1c_2^*e^{\theta_2^*+\theta_1}-e^{-\theta_2^*-\theta_1}}{\lambda_2^*-\lambda_1},
m_{22}=\frac{c_2c_2^*e^{\theta_2^*+\theta_2}-e^{-\theta_2^*-\theta_2}}{\lambda_2^*-\lambda_2},\nn\\
m_{23}=\frac{c_1c_2^*e^{\theta_2^*+\theta_1}+e^{-\theta_2^*-\theta_1}}{\lambda_2^*+\lambda_1},
m_{24}=\frac{c_2c_2^*e^{\theta_2^*+\theta_2}+e^{-\theta_2^*-\theta_2}}{\lambda_2^*+\lambda_2},\nn\\
m_{31}=\frac{c_1c_1^*e^{\theta_1^*+\theta_1}+e^{-\theta_1^*-\theta_1}}{-\lambda_1^*-\lambda_1},
m_{32}=\frac{c_2c_1^*e^{\theta_1^*+\theta_2}+e^{-\theta_1^*-\theta_2}}{-\lambda_1^*-\lambda_2},\nn\\
m_{33}=\frac{c_1c_1^*e^{\theta_1^*+\theta_1}-e^{-\theta_1^*-\theta_1}}{-\lambda_1^*+\lambda_1},
m_{34}=\frac{c_2c_1^*e^{\theta_1^*+\theta_2}-e^{-\theta_1^*-\theta_2}}{-\lambda_1^*+\lambda_2},\nn\\
m_{41}=\frac{c_1c_2^*e^{\theta_2^*+\theta_1}+e^{-\theta_2^*-\theta_1}}{-\lambda_2^*-\lambda_1},
m_{42}=\frac{c_2c_2^*e^{\theta_2^*+\theta_2}+e^{-\theta_2^*-\theta_2}}{-\lambda_2^*-\lambda_2},\nn\\
m_{43}=\frac{c_1c_2^*e^{\theta_2^*+\theta_1}-e^{-\theta_2^*-\theta_1}}{-\lambda_2^*+\lambda_1},
m_{44}=\frac{c_2c_2^*e^{\theta_2^*+\theta_2}+e^{-\theta_2^*-\theta_2}}{-\lambda_2^*+\lambda_2}.
\eeqa

Further, we can also write the obtained N-soliton solutions \eqref{5.3} into the form of the determinant ratio
\beq
r=2i\frac{det \widetilde{M}_{12}}{det M}.
\eeq
where $\widetilde{M}_{12}$ is
\eqa
\widetilde{M}_{12}=\left(\begin{array}{ll}
0&f_{1}\\
\hat{g}_{2}& M\\
\end{array}\right).\eeqa
where the matrix M is defined by Eq.\eqref{5.4}, and $f_{1}=(v_{11},v_{21}\cdots v_{2N,1}),  \hat{g}_{2}=(\hat{v}_{11},\hat{v}_{21}\cdots \hat{v}_{2N,1})^{T}$
\section{Conclusions}

In this work, by applying RH method, we establish the N-soliton solutions for the CLL-NLS equation. First of all,
the Lax pair of the coupled CLL-NLS equation are transformed to obtain the corresponding Jost solution. Then we study spectrum analysis and construct the particular RH problem, which is non-regular case. Then we derive the scattering data including the continuous scattering data $\{s_{12},s_{21}\}$ and the discrete scattering data $\{\lambda_j,\hat\lambda_j,v_j,\hat v_j\}$. With aid of reconstructing the potential, we solve the N-soliton solutions for the CLL-NLS. Finally
we obtain a simple and compact N-soliton solutions formula.

\section*{Acknowledgements}

This work is supported by the National Natural Science Foundation of China under the grant No. 11601055, Natural Science Foundation of Anhui Province under the grant No. 1408085QA06, Natural Science Research Projects of Anhui Province under Grant Nos. KJ2019A0637 and gxyq2019096.

\end{document}